\documentclass[sigconf]{acmart}

\AtBeginDocument{%
  }

\setcopyright{acmlicensed}
\copyrightyear{2018}
\acmYear{2018}
\acmDOI{XXXXXXX.XXXXXXX}

\acmConference[Conference acronym 'XX]{Make sure to enter the correct
  conference title from your rights confirmation email}{June 03--05,
  2018}{Woodstock, NY}
\acmISBN{978-1-4503-XXXX-X/18/06}

\usepackage{url}
\usepackage{color}
\usepackage{makecell}
\usepackage{subfigure}
\usepackage{colortbl}
\usepackage{caption}
\usepackage{multirow}
\usepackage{enumitem}

\usepackage{wrapfig}
\usepackage{hyperref} 
\usepackage{CJKutf8}
\usepackage{amsmath}
\usepackage{amsfonts}
\usepackage{natbib}
\usepackage{soul}

\begin{document}

\title{Finetuning Large Language Model for Personalized Ranking}


\author{Zhuoxi Bai*}
\email{bzxleon844@gmail.com}
\affiliation{
  \institution{Cashcat}
  \country{China}
}

\author{Ning Wu*}
\email{wuning@buaa.edu.cn}
\affiliation{
  \institution{Beihang University}
  \country{China}
}

\author{Fengyu Cai}
\email{fengyu.cai@tu-darmstadt.de}
\affiliation{
  \institution{Technical University of Darmstadt}
  \country{Germany}
}

\author{Xinyi Zhu}
\email{helvieluke534@gmail.com}
\affiliation{
  \institution{Shanghai University}
  \country{China}
}

\author{Yun Xiong\dag}
\email{yunx@fudan.edu.cn}
\affiliation{
  \institution{Fudan University}
  \country{China}
}

\thanks{*Both authors contributed equally to this research.~\dag The corresponding author.}

\renewcommand{\shortauthors}{Zhuoxi Bai*, Ning Wu*, Fengyu Cai, Xinyi Zhu, and Yun Xiong\dag}

\begin{abstract}

Large Language Models (LLMs) have demonstrated remarkable performance across various domains, motivating researchers to investigate their potential use in recommendation systems. However, directly applying LLMs to recommendation tasks has proven challenging due to the significant disparity between the data used for pre-training LLMs and the specific requirements of recommendation tasks. In this study, we introduce Direct Multi-Preference Optimization (DMPO), a streamlined framework designed to bridge the gap and enhance the alignment of LLMs for recommendation tasks. DMPO enhances the performance of LLM-based recommenders by simultaneously maximizing the probability of positive samples and minimizing the probability of multiple negative samples. We conducted experimental evaluations to compare DMPO against traditional recommendation methods and other LLM-based recommendation approaches. The results demonstrate that DMPO significantly improves the recommendation capabilities of LLMs across three real-world public datasets in few-shot scenarios. Additionally, the experiments indicate that DMPO exhibits superior generalization ability in cross-domain recommendations. A case study elucidates the reasons behind these consistent improvements and also underscores DMPO's potential as an explainable recommendation system. Our code and data are available at \href{https://github.com/BZX667/DMPO}{https://github.com/BZX667/DMPO}

\end{abstract}



\received{20 February 2007}
\received[revised]{12 March 2009}
\received[accepted]{5 June 2009}

\maketitle

\section{Introduction}

Recommendation systems are essential for addressing information overload and fulfilling the information needs of users \cite{guo2017deepfm, xi2023bird, xi2023towards}. Large Language Models (LLMs) have emerged as powerful tools in Natural Language Processing (NLP), demonstrating impressive capacity in natural language understanding and text generation \cite{brown2020language, touvron2023llama, wang2023recmind, zhang2023memory}. Previous studies have highlighted the rich knowledge and generalization abilities of LLMs \cite{ouyang2022training, sanh2021multitask, wei2021finetuned}. Given these strengths, LLMs are poised to revolutionize recommendation systems, particularly in few-shot or zero-shot scenarios. Consequently, researchers have begun exploring the application of LLMs in recommendation systems. However, directly applying LLMs to recommendation tasks may not yield satisfactory performance, as the pre-training data of LLMs is primarily tailored for NLP tasks, which significantly differ in purpose and requirements from recommendation tasks \cite{tallrec}.

\begin{figure*}[t]
\centering
\includegraphics[width=0.95\linewidth]{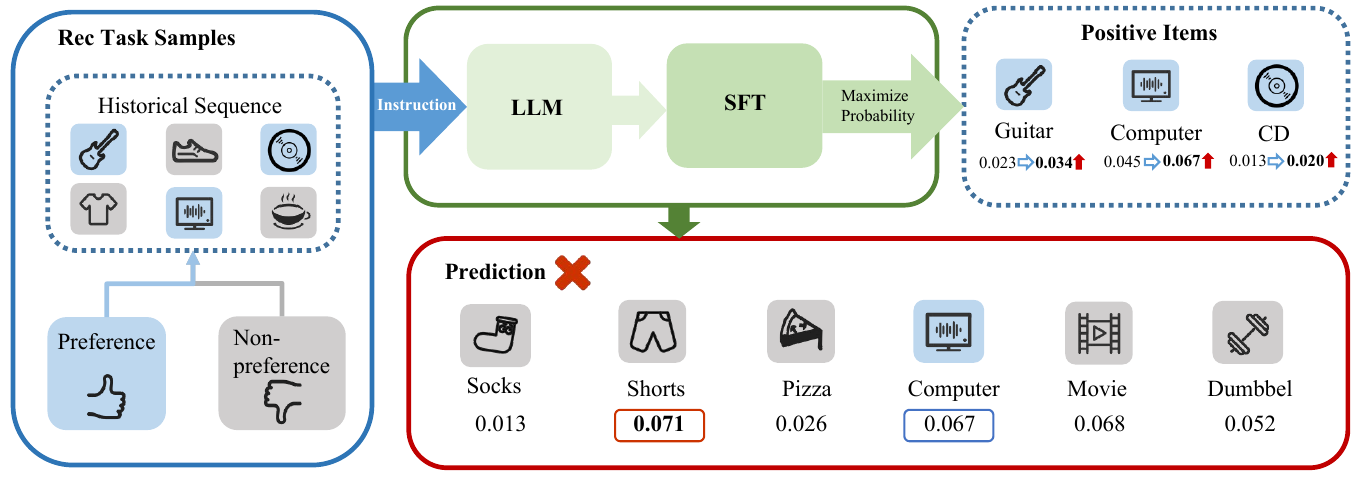}
\caption{When aligning the LLMs with recommendation tasks, applying Supervised Fine-Tuning (SFT) to LLM initially helps maximize the probability of generating each token in positive items. However, this method overlooks the potential benefits of comparing positive and multiple negative samples. As a result, although the model trained solely with SFT may improve the probability of positive items during training, it may overestimate the probability of unseen negative items during testing, leading to incorrect prediction. 
}
\label{fig:LLM_intro}
\end{figure*}

Previous work has employed in-context learning~\cite{brown2020language} to align LLMs with recommendation problems \cite{geng2022recommendation, wang2023zero}. These methods typically use closed-source models such as ChatGPT and GPT4 \footnote{https://chat.openai.com/} to rerank candidates selected by traditional recommendation models such as Matrix Factorization (MF) \cite{koren2009matrix} and LightGCN \cite{he2020lightgcn}. However, these methods only superficially incorporate LLMs, constrained by the limitations of previous methods. This is a pressing need to explore strategies for effective and efficient utilization of LLMs in recommendation tasks. Other studies, such as TALLrec \cite{tallrec}, have proposed Supervised Fine-Tuning (SFT) to bridge this gap by introducing data samples specific to recommendation tasks. TALLrec leverages the fine-tuning method used in general NLP to train on recommendation samples, aiming to achieve improved results compared to the in-context learning method. However, SFT only maximizes the probability of generating the correct answer during training. When faced with unseen incorrect answers, the model may overestimate their probability and make incorrect predictions due to the lack of negative sample learning. This approach overlooks the benefits of learning and establishing distinctions and connections between multiple pairs of positive and negative samples, as illustrated in Fig \ref{fig:LLM_intro}.

To address this limitation, we propose Direct Multi-Preference Optimization (DMPO), a streamlined framework designed to maximize the probability of generating correct answers while simultaneously minimizing the probability of generating multiple incorrect answers under a dynamic margin. The motivation for introducing DMPO arises from the constraints of the original Direct Preference Optimization (DPO) \cite{dpo}, which considered only one positive sample and one negative sample from a user's user-item interaction list, thereby restricting the diversity of training data and potential performance improvements. Drawing inspiration from contrastive learning methods that employ negative sampling strategies \cite{InstDisc, InvaSpread, MoCo, simCLR}, DMPO incorporates sampling multiple negative samples to expand the range of negative samples. This allows the model to establish more comprehensive and balanced relationships between positive and negative samples, enhancing overall performance.

To evaluate the effectiveness of DMPO, we conducted extensive experiments on three real-world public datasets: \textit{"Movielens-1M"}, \textit{"Amazon Movies and TV"}, and \textit{"Amazon Video Games"}. DMPO aims to perform binary classification by distinguishing positive items from negative items, and then rank the likelihood of a user giving a high rating to these items. When generating the answer, the model is expected to prioritize the positive items over the negative items. More details of the dataset and task can be found in Section \ref{sec:dataset} and Section \ref{sec:task introduction}. Since DMPO utilizes user-item interaction lists, it is appropriate to compare it with LLM-based methods \cite{tallrec} and traditional sequential methods\cite{GCSAN, STAMP, SASRec, GRU4Rec, BERT4Rec, SHAN, CORE}. The results demonstrate that DMPO significantly improves AUC compared to other methods in a few-shot scenario. Furthermore, in order to validate the superior generalization ability of DMPO, we conducted experiments to compare it with other LLM-based cross-domain methods \cite{tallrec} and traditional cross-domain methods \cite{BiTGCF, CMF, CLFM, DeepAPF, DTCDR}. The results highlight DMPO's breakthrough performance in cross-domain recommendation tasks. We also conducted ablation studies to investigate the factors that may affect the performance of DMPO, including the number of negative samples, the number of few-shot samples, and different base models. Additionally, we performed a case study to demonstrate the reasons why DMPO can bring improvements, and we highlighted that DMPO is an explainable recommendation system.

In total, our contributions are summarized as follows:

\begin{itemize}
    \item In order to narrow the gap between LLMs and the recommendation tasks, we propose Direct Multi-Preference Optimization (DMPO). This streamlined framework enhances LLMs' ability to model the comparative relationships between positive and negative samples for recommendation tasks.
    \item Through the comprehensive experiments, we showcase that DMPO significantly outperforms the previous LLM-based and traditional methods;
    \item Moreover, DMPO's consistent improvements across multiple domains and various types of LLMs empirically underscore its strong generalization and robustness.
\end{itemize}

\section{Related Work}
\subsection{LLM-based recommendation systems}

There have been various attempts to incorporate LLMs into recommendation systems. Despite the integration of LLMs \cite{geng2022recommendation, li2023personalized}, some efforts focus on enhancing the recommendation capability of LLMs through in-context learning methods and combining them with traditional recommendation models like MF \cite{koren2009matrix} and LightGCN \cite{he2020lightgcn}, including Chat-rec \cite{gao2023chat} and NIR \cite{wang2023zero}. However, certain approaches only depend on UserIDs and ItemIDs to represent users and items, disregarding the benefits of incorporating language information and leveraging the extensive knowledge of Language Models (LLMs) about the items. These methods might also encounter challenges in generalization due to limited training data for certain User/Item IDs \cite{hou2022towards}. Furthermore, some other works employ undisclosed models that possess initial recommendation capabilities \cite{cui2022m6}, or small models trained on large-scale downstream task data \cite{zhang2023prompt}. However, the capabilities of these methods are demonstrated to be limited by the smaller-scale models.

We have also observed some methods that utilize the SFT approach to bridge the gap between natural language processing tasks and recommendations, such as TALLRec \cite{tallrec}. TALLRec is a tuning framework for aligning LLMs with recommendations. However, solely relying on the SFT may only maximize the probability of generating the correct answer. To further enhance performance, we can introduce multiple incorrect answers into the system and establish pair-wise relationships between the correct and incorrect answers.

\subsection{Sequential Recommendation}
The user-item interaction list is utilized in DMPO, which is similar to the sequential recommendation. The sequential recommendation aims to predict the next item for a user based on their historical interactions with items \cite{fang2020deep, wang2019sequential}. Initially, Markov chain models were prevalent in sequential recommendation \cite{he2016fusing, mahmood2007learning, rendle2010factorizing, wang2015learning}. In recent times, deep learning approaches have gained popularity, including those based on RNNs \cite{cui2018mv, donkers2017sequential, GRU4Rec}, CNNs \cite{tang2018personalized, yan2019cosrec, yuan2019simple}, and attention structures \cite{BERT4Rec,kang2018self, xu2021long, zhang2019feature}. However, these models typically focus solely on UserID and ItemID, limiting their ability to generalize across different domains. Some recent studies have emphasized improving the generalization capability of sequential recommendation models through techniques like pre-training \cite{ouyang2022training, yuan2020parameter}, data augmentation \cite{qiu2022contrastive, wang2022multi, xie2022contrastive}, debiasing \cite{xie2022contrastive, wang2022unbiased, zhang2021causal, zheng2021disentangling}, and robust optimization \cite{wen2022distributionally, zhang2019feature}. However, these methods typically use UserIDs and ItemIDs to represent users and items, which can lead to insufficient training when the training data is limited. As a result, the potential of leveraging the strong generalization ability of LLMs, trained on extensive data and likely possessing rich knowledge about common items, remains largely unexplored.

\subsection{Cross-domain Recommendation}
Transfer learning aims to utilize knowledge from a source domain to enhance learning performance in the target domain or reduce the number of labeled examples needed in the target domain when achieving the same performance.  \cite{wang2022generalizing, zhuang2020comprehensive}. Cross-Domain Recommendation (CDR), inspired by transfer learning, is a promising approach to address data sparsity and the cold-start issue in the target domain by leveraging the auxiliary (source) domain. Initially, CMF \cite{singh2008relational} assumes a shared global user embedding matrix for all domains and factorizes matrices from multiple domains simultaneously. CoNet \cite{CoNet} transfers and integrates knowledge through cross-connections in feed-forward neural networks. Another set of CDR methods focuses on connecting user preferences across different domains, as demonstrated by studies such as \cite{kang2019semi, man2017cross, pan2010transfer, zhang2020learning, zhu2020deep}. These methods either utilize the user embedding learned in the source domain to initialize the user embedding in the target domain and restrict them to being closed or explicitly model the preference bridge. All the above methods are involved in exacting information from the source domain and transferring it to the target domain; however, the limitation persists because all the user/item embeddings are built on certain user/item IDs. For LLMs, using natural language provides an excellent and natural medium to transfer the features learned across domains. Consequently, the potential of using LLM-based models to accomplish the cross-domain recommendation task still remains largely unexplored.

\section{Preliminary}
In this section, we introduce preliminary knowledge of DMPO. In previous sections, we mentioned conceiving DMPO as a pair-wise ranking loss, which we consider as an important concept. We introduce what pair-wise ranking loss is and explain why DMPO shares a similar mathematical foundation with pair-wise ranking loss.

\subsection{Pair-Wise Ranking loss} 
\label{sec:pair wise loss}
Pair-wise ranking loss is a type of loss function used in supervised learning specifically for sorting problems. It calculates the loss by comparing the relative rankings of a pair of input elements. Commonly used options for this type of loss function include Hinge loss \cite{cortes1995support} and Bayesian Personalized Ranking (BPR) loss \cite{rendle2012bpr}, among others. In recommendation systems, pair-wise ranking loss is typically employed to establish the pair-wise relationship between positive items and negative items. To further illustrate this concept, the mathematical equation of BPR loss is presented in Equation \ref{eq:pair ranking loss}.

\begin{equation}
    \label{eq:pair ranking loss}
    \mathcal{L}_{BPR} = -\sum_{(u,i,j)\in\mathcal{D}}\ln\sigma(f_{u}(i) - f_{u}(j))
\end{equation}
where the $\mathcal{D}$ refers to the dataset, $u$ refers to the user, $i$ and $j$ denotes the positive and negative sample pair. $f_{u}(i)$ denotes the preference score for the user $u$ of the item $i$. Commonly used preference score functions include cosine similarity and Pearson correlation coefficient. By utilizing the BPR loss, the $f_{u}(i)$ is increased, while the $f_{u}(j)$ is decreased. Compare to the DMPO mathematical equations, which are shown in Equation \ref{eq:DMPO_multi_neg}. We found that the fundamental idea behind it is similar, except the DMPO
increases the likelihood of the preferred completions $y_w$ and decreases the likelihood of dispreferred completions $y_l$. The likelihood in the DMPO refers to the joint generating probability of multiple tokens.

\section{Methodology}
In this section, we introduce the DMPO, a streamlined framework that aims to align LLMs with recommendation tasks. 

\begin{figure*}[t]
  \centering
  \includegraphics[width=\linewidth]{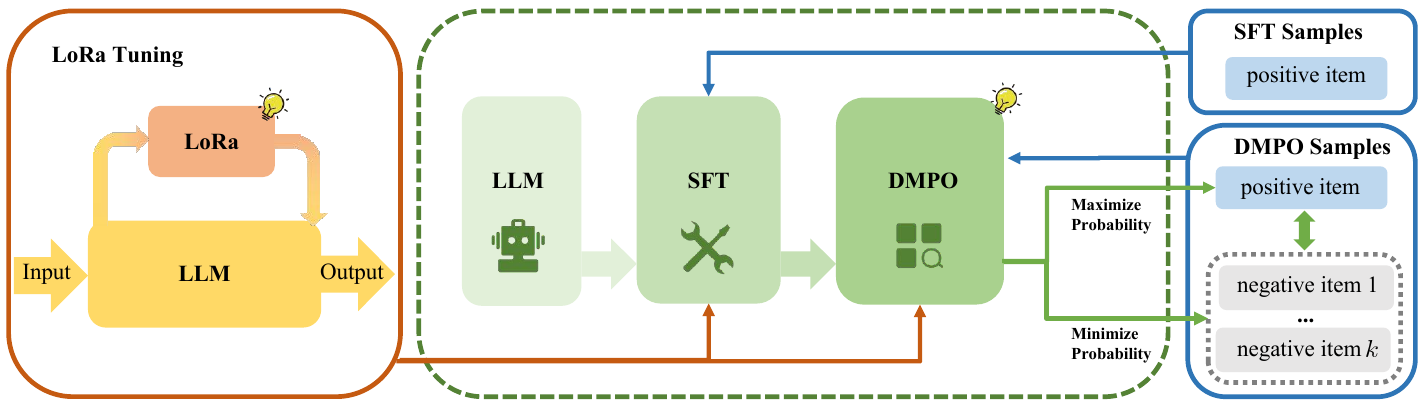}
  \caption{We first performed SFT on the base LLM model and then proceeded with DMPO. SFT samples and DMPO samples were constructed as inputs for training. Both SFT and DMPO were trained using LoRA. DMPO aims to maximize the probability of positive samples while minimizing the probability of multiple negative samples simultaneously.}
  \label{fig:DMPO framework}
\end{figure*}

\subsection{Task Introduction}
\label{sec:task introduction}
The DMPO is expected to distinguish whether the user enjoys the given items or not. The task is essentially a binary classification task. The user's item preferences history is inputted into the prompt. A pair of items from the user-item interaction list is then offered to make the prediction. One of the items, denoted as $i$, has been rated highly by the user with a score greater than 3 out of 5, making it a positive sample. The other item, denoted as $j$, has received a low rating from the user, scoring lower than 3. The model is expected to determine that the user is more likely to enjoy the positive sample item $i$ than the negative sample item $j$ by generating a recommendation that ranks the likelihood of the user giving a high rating to the items, and placing the item $i$ before the item $j$.

\subsection{Instruction Prompts}
\label{sec:DMPO}

\begin{table}
\centering
 \small
  \caption{An instruction tuning sample for DMPO. "Roman Holiday" is the positive item, and the other movies serve as negative items. The order of positive and negative samples placed in the instruction input is randomly altered across different samples to prevent the LLMs from making predictions based on the input sequence.
  }
 \label{tab:multi sample task}
 \resizebox{0.45\textwidth}{!}{
    \begin{tabular}{@{}ll@{}}
        \toprule
        \multicolumn{2}{c}{\textbf{Task Input}} \\
        \midrule
        \multirow {2}{*}{Instruction:}
         & \makecell[lp{8cm}]{You are an assistant working on movie recommendations. Here is the user's history of movies they have watched: <Waterloo Bridge>, <Rear Window>, <Forrest Gump>. Rank the likelihood of the user watching the two movies \\
         (1) <Roman Holiday> and <Iron Man> \\
         (2) <Harry Potter> and <Roman Holiday>\\
         (3) <Roman Holiday> and <Spider Man> \\
         (4) <The Lion King> and  <Roman Holiday>} \\
        \midrule
        \multicolumn{2}{c}{\textbf{Task Output}}  \\
        \midrule
        \multirow{4}{*}{DMPO Label:} 
        &  (1) ["<Roman Holiday>, <Iron Man>", "<Iron Man>, <Roman Holiday>"].  \\ 
        &  (2) ["<Roman Holiday>, <Harry Potter>", "<Harry Potter>, <Roman Holiday>"].  \\ 
        &  (3) ["<Roman Holiday>, <Spider Man>", "<Spider Man>, <Roman Holiday>"].  \\ 
        &  (4) ["<Roman Holiday>, <The Lion King>", "<The Lion King>, <Roman Holiday>"].  \\ 
        \bottomrule
  \label{table:rec-tuning}
    \end{tabular}
    }
    \vspace{-10pt}
\end{table} 

We provide an instruction sample for DMPO, as shown in Table \ref{tab:multi sample task}. Firstly, construct the \textit{Task input}, which includes the model's role settings, user-item interaction list, and positive and negative candidates. The order of positive and negative candidates in the \textit{Task input} should be randomly arranged to prevent the model from generating answers based on the input. In the user-item interaction list, only the item title is used and separated by "<>" for easier identification by the model. In the \textit{Task output}, for SFT, we only need to provide the correct answer as "<positive sample, negative sample>", with the positive sample listed before the negative sample. For DMPO, both the correct and incorrect answers should be provided in a list format "[<positive sample, negative sample>, <negative sample, positive sample>]". In the case of multiple negative sampling, multiple negative samples should be provided.



\subsection{Supervised Fine Tuning (SFT)}
SFT involves fine-tuning Language Models (LLMs) for the recommendation task using positive samples. The main objective of it is to maximize the probability of each token in the correct answers. Additionally, SFT ensures that the answers generated by LLMs adhere to the correct format, avoiding ambiguous or evasive responses. The mathematical formula is shown in equation \ref{eq:sft}.

\begin{equation}\small
\label{eq:sft}
    \max_{\Phi} \sum_{(x,y)\in\mathcal{D}} \sum_{t=1}^{|y|} \text{log} \left(  P_{\Phi}(y_{t} | x, y_{<t}) \right),
\end{equation}
where $x$ and $y$ represent the "\textit{Task Input}" and "\textit{Task Output}" in the instruction tuning data, respectively, $y_t$ is the $t$-th token of the $y$, $y_{<t}$ represents the tokens before $y_{t}$,
$\Phi$ is the original parameters of the model, and $\mathcal{D}$ is the training set. 

\subsection{Direct Multiple Preference Optimization (DMPO)}
\label{sec:DMPO tuning}
DMPO builds upon SFT by not only maximizing the probability of generating correct answers but also minimizing the probability of generating multiple negative samples. This helps the model learn and establish the subtle and complex relationships between positive and negative samples, enabling it to capture their differences and connections. Additionally, DMPO helps suppress the generation probability of key tokens in negative samples. The introduction of multiple negative sampling in DMPO expands the range of sampled negative samples compared to the original DPO\cite{dpo}, promoting a more diverse and uniform learning of negative samples. This enhancement ultimately improves the performance of the model. The mathematical formula of DMPO is shown in the equation below:
\begin{equation}
    \label{eq:DMPO_multi_neg}
    \begin{aligned}
    &\mathcal{L}_\text{DMPO}(\pi_{\theta}; \pi_{ref}) =  -\mathbb{E}_{(x, y_w, y_l)\sim \mathcal{D}}\\
    &\left[\log \sigma \left(\beta\log \frac{\pi_{\theta}(y_w\mid x)}{\pi_{ref}(y_w\mid x)}- \frac{1}{k}\sum_{i=1}^{k}(\beta \log \frac{\pi_{\theta}(y_{l\ i} \mid x_i)}{\pi_{ref}(y_{l\ i}\mid x_i))}\right)\right]
    \end{aligned}
\end{equation}

where $k$ represents the number of negative samples, $\pi_{\theta}$ represents the language model policy,  $\pi_{ref}$ denotes the reference policy, $x$ denotes the instruction inputs, $\sigma$ refers to the sigmoid activation function, $y_{w}$ and $y_{l}$ indicate the correct answers and incorrect answers that constructed as shown in Table \ref{tab:multi sample task}, $\mathcal{D}$ denotes the instruction dataset. 

The gradient of the DMPO can be written as follows:

\begin{multline*}\label{eq:gradient}
    \nabla_\theta \mathcal{L}_\text{DMPO}(\pi_\theta;\pi_{ref}) = \\
    -\beta \mathbb{E}_{(x, y_w, y_l) \sim \mathcal{D}} \bigg[ \underbrace{\sigma \left( \frac{1}{k} \sum_{i=1}^{k} (\hat{r}_\theta(x, y_l)) - \hat{r}_\theta (x, y_w) \right)}_{\text{higher weight when reward estimate is wrong}} \\
    \bigg[ \underbrace{\nabla_\theta \log \pi(y_w \mid x)}_{\text{increase likelihood of } y_w} 
    - \underbrace{\frac{1}{k} \sum_{i=1}^{k} (\nabla_\theta \log \pi(y_l \mid x))}_{\text{decrease likelihood of } y_l} \bigg] \bigg],
\end{multline*}

where $k$ also refers to the number of negative samples. $\hat{r}_\theta(x, y) = \beta \log \frac{\pi_\theta(y \mid x)}{\pi_{ref}(y \mid x)}$ denotes the reward implicitly defined by the language model $\pi_\theta$ and reference model $\pi_{ref}$. The mathematical formulation of DMPO shows that instead of computing the probability of generating negative samples $\hat{r}_\theta(x, y_{l})$ using a single negative item, it is now computed as the average value of $k$ negative samples. It promotes a more diverse and uniform learning of negative samples, which enhances the performance in recommendation tasks.

\subsection{Prediction}
\label{prediction}
To predict the generation probabilities of positive and negative candidates, we calculate the probability of each token in each candidate and obtain their average, denoted as $P$. We then use this average probability $P$ to calculate the evaluation metric AUC. The mathematical formula of $P$ is shown in the equation below:

\begin{equation}
    \label{eq:prediction}
    \begin{aligned}
    &P = \frac{\pi_{\theta}(y\mid x)}{n_y},
    \end{aligned}
\end{equation}
where $n_y$ represents the number of tokens in the candidate $y$ that needs to be predicted, $\pi_{\theta}$ represents the model we have established, $x$ denotes the instruction inputs.

\section{Experiments}

\subsection{Dataset}
\label{sec:dataset}
We conduct experiments on three real-world public datasets, which are \textit{MovieLens 1M}, \textit{Amazon Movies and TV}, and \textit{Amazon Video Games}. All three datasets contain a list of user-item interactions, with ratings of products. In our experiment, following the prior studies \cite{he2020lightgcn, zhang2023reformulating}, we categorize the ratings into two classes: ratings $<=3$ indicate users who do not prefer the movie (negative item), while ratings $>3$ indicate users who prefer the movie (positive item). Data filtering is performed based on the number of user-item interactions, ensuring there are a minimum of $5$ positive and $5$ negative samples, with the length of the interaction list limited to $40$ items. The experiment is conducted in a few-shot learning scenario. For training, validation, and testing, we randomly selected $100$, $100$, and $1000$ samples, respectively. Statistics and more details can be found in our released data. 

\subsubsection{MovieLens 1M} It is a movie dataset commonly utilized in recommendation tasks, collected by the GroupLens Research Project at the University of Minnesota \footnote{https://grouplens.org/datasets/movielens/1m/}. The dataset comprises $1,000,209$ anonymous ratings for around $3,900$ movies provided by $6,040$ MovieLens users who joined the platform in 2000. Each user has rated a minimum of 20 movies. Data collection took place on the MovieLens website over seven months from September 19th, 1997, to April 22nd, 1998. User ratings range from $(1-5)$.

\subsubsection{Amazon Datasets} This dataset contains 142.8 million reviews and metadata of Amazon products from May 1996 to July 2014 \footnote{https://nijianmo.github.io/amazon/index.html}. It includes subsets like Books, Movies and TV, Video Games, Electronics, and more. For our study, we chose the "\textit{Amazon Movies and TV}" and "\textit{Amazon Video Games}" datasets, which include reviewer ID, item ID, item title, item brand, user-item interactions, ratings, and user reviews. In our experiment, we only use the item title and the user-item interaction history for DMPO to make the prediction.

\begin{table*}[t]
\setlength{\abovecaptionskip}{0cm}
\setlength{\belowcaptionskip}{0cm}
\caption{
Performance comparison among traditional sequential recommendation baselines, LLM-based methods, and DMPO. The reported results are presented as the AUC multiplied by 100, with boldface indicating the highest score. $\ddagger$: significantly better than all baselines with a t-test $p$<0.01.
}
\setlength{\tabcolsep}{4mm}{
\resizebox{\textwidth}{!}{
\begin{tabular}{ccccccccccc}
\toprule
Dataset &GCSAN  &STAMP &SASRec & GRU4Rec   & BERT4Rec  & SHAN  & CORE & TALLrec & OurMethod \\
 \midrule
 \multirow{1}{*}{MovieLens 1M}& 52.77  & 53.75& 54.77& 52.75  & 47.72  & 51.18    &50.32 & 67.79 & \textbf{70.20}$\ddagger$\\
\midrule
 \multirow{1}{*}{Amazon Movie}& 51.07& 53.04& 49.05& 47.51 & 50.17  & 52.88    &49.55 & 57.62 & \textbf{66.60}$\ddagger$\\
\midrule
\multirow{1}{*}{Amazon Game}& 52.11& 51.27& 50.23  & 51.26   & 50.81   & 50.35   & 49.96 & 53.17 &  \textbf{64.51}$\ddagger$\\
\bottomrule
\end{tabular}
}}

\label{tab:over_all_compare}
\end{table*}

\subsection{Baseline}
We compared DMPO with both LLM-based and traditional sequential recommendation methods. To ensure a fair comparison, we followed the same setup as DMPO for the LLM-based method, including tuning instructions and training, validation, and test data. For the traditional sequential recommendation methods, we generated train, valid, and test datasets using the same data as DMPO, which include item ID, user ID, and labels derived from user ratings.

\subsubsection{LLM-based methods}
We conducted an experiment to compare DMPO with previous work that employed LLM for the recommendation task. \textbf{(i) TALLRec} \cite{tallrec} is an LLM-based approach that aims to bridge the gap between training tasks for LLMs and recommendation tasks. It incorporates recommendation data into LLMs through rec-tuning, which involves constructing instruction-tuning samples using recommendation data and conducting SFT. We use the implementation provided by the authors.\footnote{\url{https://github.com/SAI990323/TALLRec}.}

\subsubsection{Traditional methods}
\label{sec:overall_comnpare_trad_baseline}
Since DMPO utilizes the user's preference item list to construct the prompt, it bears similarities to sequential recommendation systems. Therefore, we experimented to compare DMPO with the traditional sequential methods. \textbf{(i) GRU4Rec~\cite{GRU4Rec}} is an RNN-based sequential recommender, which utilizes GRU to encode historical interactions.
\textbf{(ii) GCSAN~\cite{GCSAN}} focuses on addressing session recommendation problems by utilizing Graph Neural Networks (GNNs) and self-attention networks.
\textbf{(iii) SASRec \cite{SASRec}} is a classic transformer-based sequential recommender. 
\textbf{(iv) STAMP \cite{STAMP}} is a short-term attention/memory priority model. It combines long-term memory to capture general interests and short-term memory to account for current user interests
\textbf{(v) BERT4Rec\cite{BERT4Rec}} is a sequential recommendation model that utilizes deep bidirectional self-attention to model user behavior sequences. It addresses the limitations of unidirectional models by allowing each item in the sequence to fuse information from both left and right sides.
\textbf{(vi) SHAN\cite{SHAN}} is a two-layer hierarchical attention network that considers user long-term preferences and dynamic characteristics. It incorporates both user-item and item-item interactions in a non-linear way.
\textbf{(vii) CORE\cite{CORE}} is a recommender that addresses the inconsistent prediction issue in the session-based recommendation by unifying the representation space for session embedding and item embeddings. It achieves this through a representation-consistent encoder and a robust distance-measuring method.

\begin{figure*}[t]
  \centering
  \includegraphics[width=1.0\textwidth]{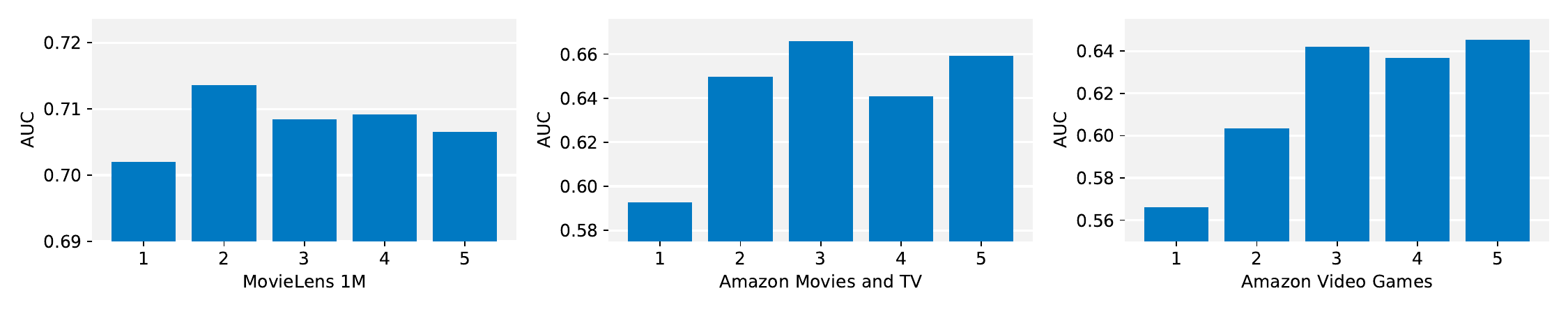}
  \caption{Performance for different numbers of negative samples in DMPO. The x-axis label represents the number of negative samples. }
  \label{fig:multi neg sample}
  \vspace{-10pt}
\end{figure*}

\subsection{Evaluation Metric}
Although the task's output format involves ranking the likelihood of user preference for items, the task itself is fundamentally a binary classification problem. The model is tasked with distinguishing between the positive and negative samples and placing the positive sample ahead of the negative sample. The objective is to differentiate between positive and negative samples and establish pair-wise relationships between correct and incorrect answers. Therefore, we utilize a widely used evaluation metric in recommendation systems: the \textit{Area Under the Receiver Operating Characteristic (AUC)}.

\subsection{Implementation Details}
\label{Implementation Details}
To ensure sequence lengths, we truncated the user-item interaction list to 40, while ensuring a minimum of 5 positive and 5 negative items. For training, validation, and testing, we randomly selected 100, 100, and 1000 samples, respectively. For most experiments, we utilized \textit{Llama-2-7b-chat-hf} \footnote{https://huggingface.co/meta-llama/Llama-2-7b-chat-hf} as the base model. The training of LoRA \cite{lora} was implemented using Llama Factory \cite{zheng2024llamafactory}, which is an efficient LLM tuning framework. We examined the learning rates for all methods in the set {1e-3, 1e-4, 1e-5, 1e-6} to identify the optimal choice and adopted a cosine decay policy for the learning rate. For LoRA rank, we conducted experiments with {8, 16, 32} to determine the optimal choice, using q, k, and v as LoRA targets. In the case of the traditional method, we utilized the implementation provided by RecBole\footnote{\url{https://github.com/RUCAIBox/RecBole}}, a project that replicates and enhances recommendation algorithms within a unified framework. Subsequently, we conducted five runs using different random seeds and presented the averaged results. Further details are provided in the code.

\section{Analysis}
In this section, we conduct analysis of our experiments to address the following research questions:

\begin{itemize}[leftmargin=*]
    \item [-] \textbf{RQ1:} 
    How does DMPO perform compared to current LLM-based and traditional sequential recommendation models? 
    \item [-] \textbf{RQ2:} 
    What factors affect the performance of DMPO?
    \item [-] \textbf{RQ3:} 
    What about the generalization ability of DMPO? How does it perform under cross-domain recommendation?
        \item [-] \textbf{RQ4:} 
    Why does DMPO lead to improvements?
\end{itemize}

\subsection{Performance Comparison (RQ1)}
\label{sec:performance comparison overall}
We conducted the experiment in a few-shot learning scenario to compare the recommendation performance of different methods. The results are presented in Table \ref{tab:over_all_compare}. From the table, we draw the following observations: 1) Our method significantly outperforms both traditional sequential recommendation methods and LLM-based methods, demonstrating the superiority of the DMPO framework. 2) The LLM-based method shows improvements compared to the traditional sequential recommendation method, aligning with expectations but still lower than the DMPO. 3) The traditional sequential recommendation method performs similarly to random guessing ($AUC\approx0.5$) in this scenario, indicating that its effectiveness is limited by the amount of training data and its inability to quickly learn recommendation capabilities with a small number of training samples.

\subsection{Ablation Study (RQ2)}

\subsubsection{Number of Negative samples in DMPO}
\label{sec:DMPO num}
In the DMPO, multiple negative sampling is introduced into the framework, as discussed in section \ref{sec:DMPO}. One important question that arises is the optimal number of negative samples to include in DMPO and how this quantity affects performance. To address this question, we conducted ablation studies with varying numbers of negative samples denoted as $k$. The results are shown in Figure \ref{fig:multi neg sample}. Based on the results, we make the following observations: 1) Increasing the number of negative samples in DMPO improves the performance compared to using only a single negative sample. 2) The performance tends to stabilize when the number of negative samples ranges from three to five. This improvement is derived from the introduction of more diverse negative samples for comparison. The multiple negative sampling enhances the diversity and uniformity of negative samples, enabling the model to establish more comprehensive and rich comparisons between positive and negative samples. However, the extent of this improvement is not infinite, and adding more negative samples beyond five may lead to additional computational overhead, reducing the efficiency of the model, and providing limited additional benefits. Moreover, the number of negative samples is also limited by the total length of the user-item interaction list. Therefore, we did not further increase the number of negative samples.

\subsubsection{DMPO and SFT with Varying Few-Shot Samples}

\begin{figure}[t!]
\centering
\begin{minipage}[b]{1\linewidth}
\centering
\includegraphics[width=\linewidth]{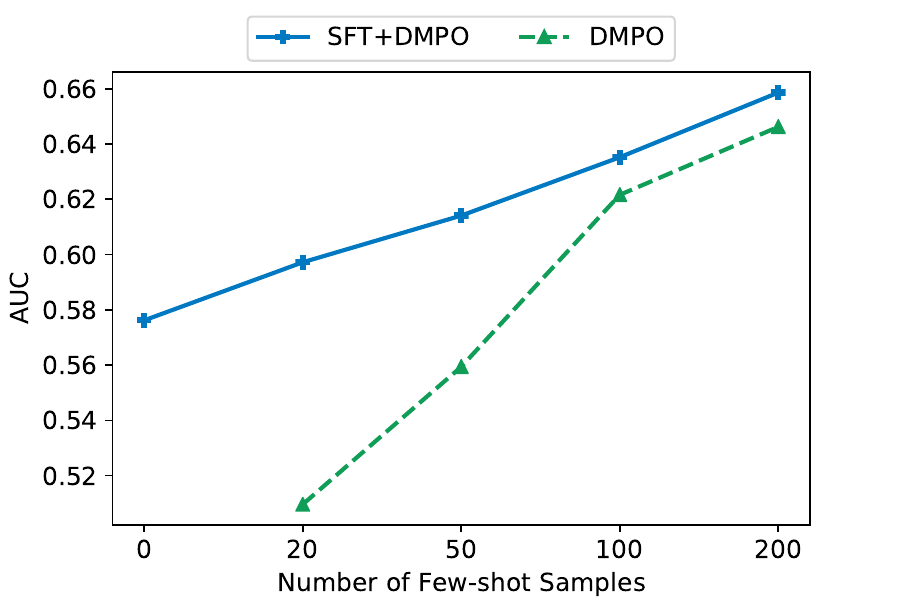}\vspace{-0.2cm}
\end{minipage}
\caption{Performance of using SFT+DMPO  and using only DMPO is compared under different few-shot sample numbers.}\label{fig:few_shot_number}\vspace{-0.3cm}
\vspace{-5pt}
\end{figure} 

We conducted experiments on the \textit{Amazon Movies and TV} dataset to investigate two questions. The first question was about the difference in performance between using DMPO alone and using SFT+DMPO, and to clarify the roles of SFT and DPO. The second question was about the impact of different numbers of few-shot samples on the model. The experimental results are presented in Figure \ref{fig:few_shot_number}. The observation can be drawn as follows: 1) When the number of few-shot samples is less than or equal to 50, using DMPO alone yields lower results than using SFT+DMPO. However, when the number of few-shot samples increases to 100 or more, the performance gap between DMPO and SFT+DMPO is reduced. The performance of both approaches is similar, with a difference of only about 1.3 percent. This suggests that DMPO plays an important role in improving model performance. 2) As the number of few-shot samples increases, the performance of both DMPO and SFT+DMPO shows a clear upward trend. However, with only 20 few-shot samples, the AUC can reach around 87 percent of the AUC achieved with 200 few-shot samples. This demonstrates the high efficiency of DMPO.

\subsubsection{Generalization across Base Models}

\begin{figure}[t!]
\centering
\begin{minipage}[b]{1\linewidth}
\centering
\includegraphics[width=\linewidth]{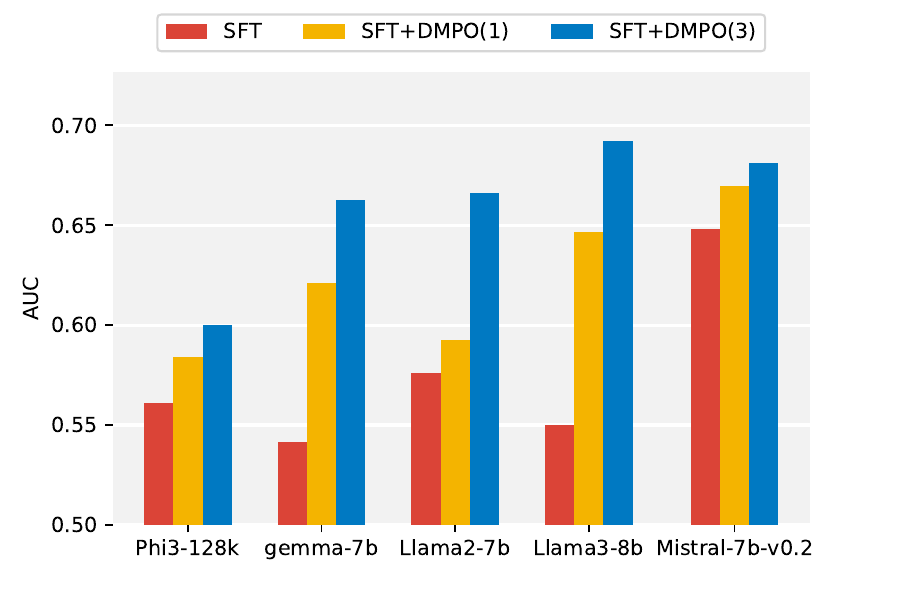}\vspace{-0.2cm}
\end{minipage}
\caption{Performance for different base models of DMPO. In the label "SFT+DMPO(1)" and the label "SFT+DMPO(3)", the number in the brackets refers to the number of negative samples used in DMPO. }\label{fig:multi_base_model}\vspace{-0.3cm}
\end{figure}

\begin{table*}[t]
\setlength{\abovecaptionskip}{0cm}
\setlength{\belowcaptionskip}{0cm}
\caption{
 Comparison of cross-domain generalization performance between DMPO, traditional cross-domain recommendation models, and LLM-based models under few-shot scenarios. The reported results are presented as the AUC multiplied by 100, with boldface indicating the highest score, $\ddagger$, which is significantly better than all baselines according to a t-test with $p$<0.01. "Movie to Game" means the "\textit{Amazon Movies and TV}" is the source domain and the "\textit{Amazon Video Games}" is the target domain. "Game to Movie" means that "\textit{Amazon Video Games}" is the source domain and "\textit{Amazon Movies and TV}" is the target domain. 
}
\setlength{\tabcolsep}{4mm}{
\resizebox{\textwidth}{!}{
\begin{tabular}{ccccccccc}
\toprule
Dataset &BiTGCF &CMF  &DTCDR &CoNet & CLFM   & DeepAPF  & TALLRec & DMPO \\
 \midrule
 \multirow{1}{*}{Movie to Game}&52.57 & 49.16 &48.29 & 49.7  & 51.25    &50.49 & 54.61 & \textbf{62.73}$\ddagger$\\
\midrule
 \multirow{1}{*}{Game to Movie} &52.4 &52.70 & 48.73  & 47.91    &53.33 &49.13 & 52.59 & \textbf{64.55}$\ddagger$\\
\bottomrule
\end{tabular}
}}

\label{tab:cross-domain compare}
\end{table*}

Recently, many researchers in the field of LLMs have publicly released their models and model parameters, including \textit{Llama-3-8B} \footnote{https://huggingface.co/meta-llama/Meta-Llama-3-8B}, \textit{gemma-7b} \footnote{https://huggingface.co/google/gemma-7b}, 
\textit{Phi-3-mini-128k-instruct} \footnote{https://huggingface.co/microsoft/Phi-3-mini-128k-instruct} and \textit{Mistral-7B-Instruct-v0.2} \footnote{https://huggingface.co/mistralai/Mistral-7B-Instruct-v0.2}. We have a strong interest in evaluating the performance of our method on various open-source language model-based recommendation systems (LLMs), particularly those models that claim to have SOTA performance. As a result, we conducted experiments on multiple recently released LLMs using the \textit{Amazon Movies and TV}. To validate the effectiveness of DMPO, we conducted experiments under three different conditions: (1) SFT; (2) SFT and DMPO using only one negative sample; (3) SFT and DMPO using three negative samples. In Section \ref{sec:DMPO num}, we found that employing three negative samples in the DMPO yielded relatively stable improvements.

The results, shown in Figure \ref{fig:multi_base_model}, demonstrate the following conclusions: 1) Firstly, on various base models of LLMs, performing DMPO can lead to an increase in AUC cross different base models; 2) The increasing number of negative sample brings further performance improvements; 3) Among all models, the \textit{Mistral-7B-Instruct-v0.2} shows the smallest improvement after utilizing DMPO, while the \textit{Llama-3-8B} model shows the largest improvement, which may be related to the structures, train methods, and parameter choices of the different base models.

\subsection{Cross-Domain Analysis (RQ3)}
To investigate the generalization ability of DMPO, we conducted cross-domain experiments between the "\textit{Amazon Movies and TV}" and the "\textit{Amazon Video Games}". Initially, we used "\textit{Amazon Movies and TV}" as the source domain and "\textit{Amazon Video Games}" as the target domain. Subsequently, we swapped the roles of the source and target domains for further analysis. In the cross-domain evaluation, the data from the source domain was used for training, while the data from the target domain was used for validation and testing. The datasets remained consistent with the original datasets described in Section \ref{sec:dataset}.

\subsubsection{Baselines}
We compared DMPO with the traditional cross-domain method and LLM-based cross-domain method. For the traditional cross-domain approach, we considered the following methods: \textbf{(i) CMF} \cite{CMF}, a well-known cross-domain recommender using Collective Matrix Factorization. \textbf{(ii) CLFM} \cite{CLFM}, a cross-domain recommender that identifies shared and unique rating patterns between domains, focusing solely on shared patterns for knowledge transfer. \textbf{(iii) DTCDR} \cite{DTCDR}, a cross-domain model employing an adaptive embedding-sharing strategy based on multi-task learning to enhance recommendation performance on both richer and sparser domains simultaneously. \textbf{(iv) DeepAPF} \cite{DeepAPF}, a cross-domain recommender that captures cross-domain common interests and domain-specific interests using an attention network. \textbf{(v) BiTGCF} \cite{BiTGCF}, a cross-domain recommender based on a graph collaborative filtering network, integrating common user features with target domain-specific features.

\subsubsection{Performance Comparsion}
The results presented in Table \ref{tab:cross-domain compare} indicate the following findings, 1) DMPO demonstrates a significant improvement compared to both LLM-based and traditional cross-domain methods. 2) The cross-domain recommendation result for DMPO was slightly lower compared to training on the target domain data, which is aligned with the exception. 

\begin{figure}[t!]
\centering
\begin{minipage}[b]{1\linewidth}
\centering
\includegraphics[width=\linewidth]{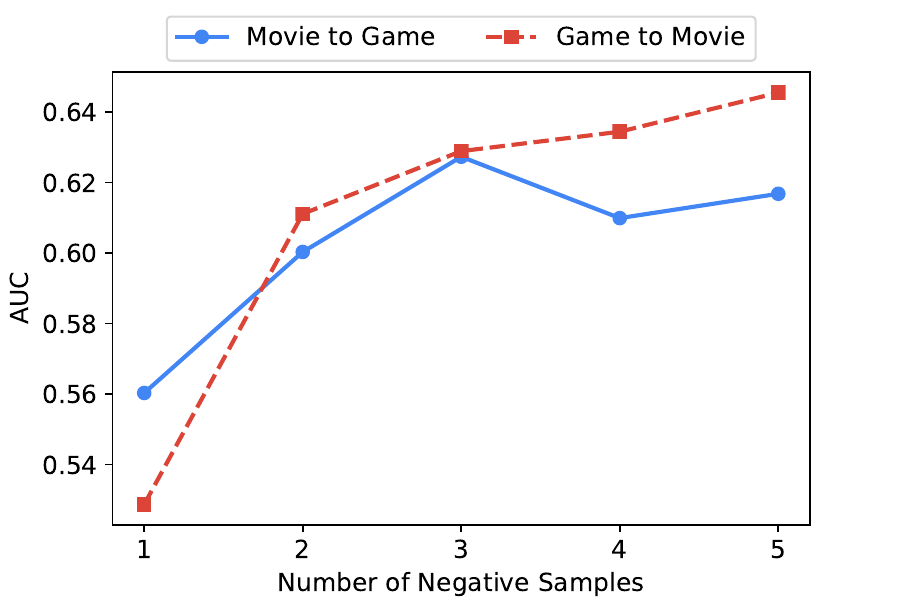}\vspace{-0.2cm}
\end{minipage}
\caption{Performance of the DMPO in cross-domain recommendation varies with different numbers of negative samples. The x-axis represents the number of negative samples}\label{fig:cross_domain_compare}\vspace{-0.3cm}
\end{figure}

We also conduct the experiment to investigate the influence of varying numbers of negative samples in DMPO in cross-domain recommendation. The results present in Figure \ref{fig:cross_domain_compare} indicate the following observations.  1) The increasing number of negative samples enhances the performance in the cross-domain context. 2) The performance tends to stabilize when the number of negative samples reaches three to five.

\begin{figure}[t!]
\centering
\begin{minipage}[b]{\linewidth}
\centering
\includegraphics[width=\linewidth]{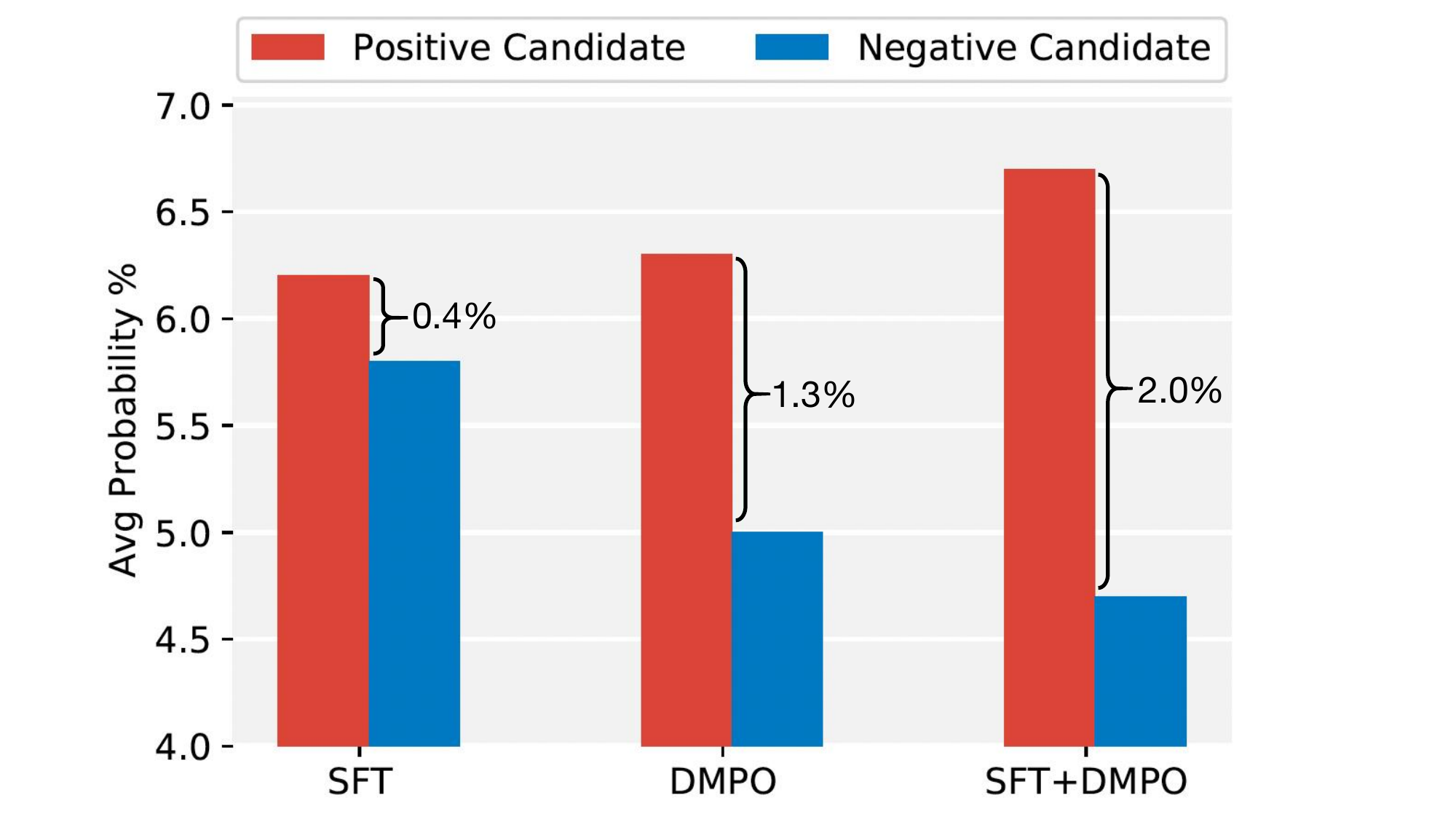}\vspace{-0.2cm}
\end{minipage}
\caption{Average conditional probabilities for positive and negative candidates of "\textit{Movielens-1M}" test dataset. Predictions of the model using only SFT, the model using only DMPO, and the model using SFT+DMPO and the gap of probabilities between candidates are presented.}\label{fig:case study gap}\vspace{-0.3cm}
\vspace{-5pt}
\end{figure}

\subsection{Case Study (RQ4)}

\begin{figure*}[t]
  \centering
  \includegraphics[width=1.0\textwidth]{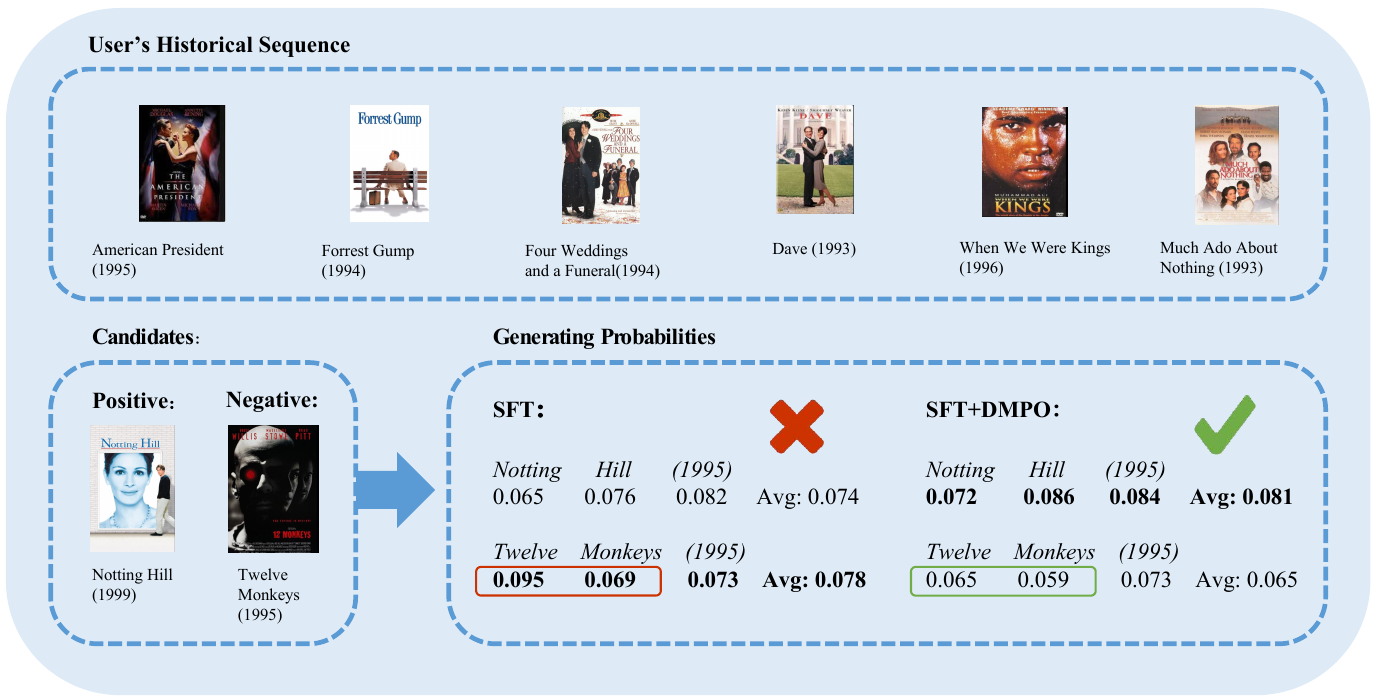}
  \caption{A case study sample illustrates the reason why DMPO can improve the performance. The user's historical sequence, positive and negative candidates, and the conditional probabilities for generating each token are provided. model using only SFT made an incorrect prediction, while model using SFT+DMPO made the correct prediction. }
  \label{fig:case study sample}
\end{figure*}

To investigate why DMPO leads to improvements, we conducted a case study. By inputting a user's historical sequence, positive and negative candidates, we obtained the conditional probability of each token of candidates. Through a specific example, as shown in Figure \ref{fig:case study sample}, we made the following observations: 1) the gap between positive and negative candidates increased after applying DMPO; 2) the conditional probability of key tokens in negative candidates was suppressed after DMPO, while this was not observed in the model that only underwent SFT. This suggests that through DMPO, the model achieved better discrimination between positive and negative candidates. 

The average conditional probability data for the "\textit{Movielens-1M}" test dataset also supports this conclusion, as shown in Figure \ref{fig:case study gap}. The gap between positive and negative candidates' conditional probabilities increased from 0.4 percent in the model using only SFT to 2.0 percent in the model using SFT+DMPO, and the average conditional probabilities of negative candidates decreased. This further validates our previous findings. It is worth noting that the above conclusion can also be obtained by using DMPO alone, as shown in the label \textit{DMPO}, which means that DMPO can provide performance benefits even when used independently.

\subsection{Explainable Recommendation System}
The explainable recommendation system can improve the acceptance of recommended products, persuade users to purchase them, and even enhance the overall trust of the system \cite{herlocker2000explaining}. The DMPO can provide interpretability for recommendation tasks \cite{rafailov2024r}. As shown in Section \ref{prediction}, the model calculates the probability of generating each token in the candidates and assigns higher probabilities to the important tokens, indicating that the recommendation is made based on those crucial tokens. This explains why certain candidates are chosen for item recommendations.

\section{Conclusion}

In this study, we introduced Direct Multiple Preference Optimization (DMPO), a streamlined framework that effectively aligns LLMs with recommendation tasks. DMPO can be seen as a pair-wise ranking loss that distinguishes between positive and negative samples. Through multiple negative sampling, DMPO enhances the performance of LLM-based recommenders by maximizing the probability of positive samples and minimizing the probability of multiple negative samples simultaneously. This enhances the diversity and uniformity of negative samples, enabling the model to learn and establish more comprehensive and accurate positive-negative sample relationships. Through extensive experiments, we compared the performance of DMPO with other methods, including LLM-based and traditional sequential recommendation approaches. Significant improvements were observed with DMPO in few-shot scenarios. Additionally, we evaluated the generalization ability of DMPO in cross-domain recommendation tasks, comparing it with traditional cross-domain methods and LLM-based cross-domain methods. DMPO exhibited superior performance in cross-domain recommendation. Ablation studies were conducted to analyze the factors contributing to the enhancements in DMPO, and case studies explored the reasons for the improvements. Moreover, DMPO is also highlighted as an explainable recommendation system. Moving forward, our future research will focus on exploring more efficient methods to leverage the recommendation capabilities of LLMs.

\nocite{*}
\bibliographystyle{ACM-Reference-Format}
\bibliography{reference}

\end{document}